# HABITABILITY OF EXOPLANETARY SYSTEMS WITH PLANETS OBSERVED IN TRANSIT

BARRIE W. JONES AND P. NICK SLEEP

**Astronomy Group, The Open University, Milton Keynes, MK7 6AA, UK**

**b.w.jones@open.ac.uk**

# ABSTRACT

We have used the *measured* properties of the stars in the 79 exoplanetary systems with one or more planets that have been observed in transit, to estimate each system's *present* habitability. Such systems have the advantage that the inclination of the planetary orbits is known, and therefore the actual mass of the planet can be obtained, rather than the minimum mass in the many systems that have been observed only with the radial velocity technique. The measured stellar properties have been used to determine the present location of the classical habitable zone (HZ). To establish habitability we use the estimated distances from the giant planet(s) within which an Earth-like planet would be inside the gravitational reach of the giant. These distances are given by $nR_H$, where $R_H$ is the Hill radius of the giant planet and $n$ is a multiplier that depends on the giant's orbital eccentricity $e_G$ and on whether the orbit of the Earth-like planet is interior or exterior to the giant planet. We obtained $n_{int}(e_G)$ and $n_{ext}(e_G)$ in earlier work and summarize those results here. We then evaluated the present habitability of each exoplanetary system by examining the penetration of the giant planet(s) gravitational reach into the HZ. Of the 79 transiting systems known in April 2010, only 2 do *not* offer safe havens to Earth-like planets in the HZ, and thus could not support life today. We have also estimated whether habitability is possible for 1.7 Gyr into the past i.e. 0.7 Gyr for a heavy bombardment, plus 1.0 Gyr for life to emerge and thus be present today. We find that, for the best estimate of each stellar age, an additional 28 systems do *not* offer such sustained habitability. If we reduce 1.7 Gyr to 1.0 Gyr this number falls to 22. However, if giant planets orbiting closer to the star than the inner boundary of the HZ, have got there by migration through the HZ, and if this ruled out the subsequent formation of Earth-like planets, then, of course, none of the presently known transiting exoplanetary systems offers habitability. Fortunately, this bleak conclusion could well be wrong.

As well as obtaining results on the 79 transiting systems, this paper demonstrates a method for determining the habitability of the cornucopia of such systems that will surely be discovered over the next few years.

# 1 INTRODUCTION

An exoplanet is detected in transit through the periodic reductions in the apparent brightness of its star, when, once each orbit, the planet passes between us and the star. The planet itself is not observed directly. Because the larger the planet the easier it is to detect in transit, there is an observational bias towards discovering larger planets. The masses of the transiting planets have been measured from a combination of the fact of transit and radial velocity (RV) measurements. Jones (2004, Chapters 9 and 10) describes the various techniques for finding exoplanets.

There are 79 known transiting systems (Schneider[1] April 2010), three with two planets, making 82 transiting planets in total. It is instructive to compare their radii, masses, and densities with the values for the planets in the Solar System. Table 1 gives the Solar System data. The inner four planets have densities that indicate a silicate-iron composition for the bulk of the planet. The densities of Jupiter and Saturn indicate the predominance hydrogen and helium, those of Uranus and Neptune indicate abundant hydrogen and helium but with a greater proportion of water than in Jupiter and Saturn. There is plenty of other evidence for these broad compositions. A type of planet not found in the Solar System is one rich in carbon compounds. Another absence is of planets much richer in water than the Earth. Water-rich planets could well have radii and masses between that of the Earth, and that of Uranus or Neptune (Marcy 2009).

TABLE 1 PROPERTIES OF THE PLANETS IN THE SOLAR SYSTEM (IN OUTWARD ORDER)

| | Mercury | Venus | Earth | Mars | Jupiter | Saturn | Uranus | Neptune |
|---|---|---|---|---|---|---|---|---|
| radius/$r_E$ | 0.3826 | 0.9489 | 1 | 0.5325 | 11.21 | 4.725 | 2.004 | 1.941 |
| mass/$m_E$ | 0.0553 | 0.8150 | 1 | 0.1074 | 317.8 | 95.16 | 14.54 | 17.15 |
| mean density/(kg $m^{-3}$) | 5430 | 5240 | 5520 | 3350 | 1330 | 690 | 1270 | 1640 |

$r_E$ denotes the radius of the Earth, 6378 km, and $m_E$ the mass of the Earth, 5.9722 × $10^{24}$ kg

Among the known transiting systems only three planets have masses less than that of Uranus. These are the two planets of CoRoT-7, and the one planet of GJ 1214. Table 2 lists their radii, masses, and densities, and those of Uranus and the Earth. All three transiting planets have masses several times that of the Earth. CoRoT-7 b has a mean density close to the Earth's, indicating a silicate-iron composition, possibly enriched in carbon compounds or water compared to the Earth (Queloz et al. 2009). The mean density of GJ 1214 b is sufficiently low to indicate a considerable water component (Charbonneau et al. 2009). The rest have densities indicating that they are dominated by hydrogen and helium, with water constituting a greater mass fraction as the mass declines.

[1] [1] [1] Schneider J., http://exoplanet.eu/

TABLE 2 PROPERTIES OF THE THREE TRANSITING PLANETS THAT HAVE MASSES LESS THAN URANUS

| | CoRoT-7 b | CoRoT-7 c | GJ 1214 b | Uranus | Earth |
|---|---|---|---|---|---|
| radius/$r_E$ | 1.68 | ??? | 2.71 | 2.004 | 1 |
| mass/$m_E$ | 4.80 | 8.39 | 5.69 | 14.54 | 1 |
| mean density/(kg $m^{-3}$) | 5460 | ??? | 1575 | 1270 | 5520 |

Among the remaining transiting planets only six are less massive than Saturn and thus the very great majority are giant planets. Therefore, an important conclusion is that in nearly all of the transiting systems there could be as yet undiscovered planets with masses within a factor of a few of the Earth's mass. The possibility of non-discovery is considerably increased if the planets are in larger orbits than those of the population of currently known transiting exoplanets, where in only five cases is the planet in an orbit with a semimajor axis greater than 0.1 AU: the largest is the 1.186 AU orbit of the massive planet HAT-P-13 c. The greater the size of the orbit, the less likely that the planet will transit its star. Discovery of not-quite-transiting planets would then rely on RV measurements of systems drawn to our attention by their having at least one transiting planet, a daunting prospect for a low mass planet in a relatively large orbit, because of the relatively small radial velocity induced in the star by such a planet.

A concept central to our work is the *Earth-like planet*, by which we mean that the planet has a mass $m$ within a factor of a few of the Earth, and a broadly similar composition. This means that its density $\rho$ will not differ greatly from the Earth's, and therefore that its radius, proportional to $(m/\rho)^{1/3}$, is tightly constrained. Such a planet is also likely to have an atmosphere.

We have therefore used computer simulations to see whether Earth-like planets could be dynamically stable, and in particular remain *confined* to the classical habitable zone of each system for long enough for life to emerge. If so, then it is possible that life is present on any such planets.

It is the possibility of life being present that drew us to the work reported here. The second attraction was that for planets in transit we know the angle at which the orbit is presented to us, close to edgewise, and so the mass obtained from follow-up RV measurements is essentially the actual mass of the planet, and not the minimum mass that is obtained from RV measurements alone (Jones et al. 2006).

As well as obtaining results on the 79 transiting systems, this paper demonstrates a method for determining the habitability of the many such systems that will surely be discovered over the next few years, doubtless with planets in larger orbits, increasing the proportion of systems where an Earth-like planet could remain confined to the classical habitable zone.

### *1.1 The essence of our approach*

The essence of our approach is to establish the location of the classical habitable zone around a star known to have one or more giant planets, and to determine the gravitational reach of each giant. If the gravitational reach traverses the classical habitable zone, confinement of an Earth-like planet is ruled out and the system is classified as uninhabitable. If all, or a significant proportion of the classical habitable zone, is free from penetration, then confinement is possible over some or all of the classical habitable zone and the system is classified as habitable. As a star ages, the classical habitable zone migrates outwards and so there is a distinction between instantaneous habitability and habitability for long enough that life could emerge. This we take to be 1.7 Gyr, including the first 0.7 Gyr on the main-sequence to allow for a presumed heavy bombardment (Jones 2004, Chapter 3).

To locate the classical habitable zone we need the luminosity $L$ of the star and its effective temperature $T_e$ (see Section 2.1). We have used the measured properties of the stars in the known transiting systems. Clearly this yields their instantaneous habitability *today*. $T_e$ is obtained from the measured properties spectral type and luminosity class. To obtain $L$, we require the stellar distance $d$, the apparent visual magnitude $V$, and the bolometric correction $BC$. $BC$ depends on the spectral type and luminosity class. See Section 2.1 for details. We can also estimate, from the stellar age, whether life has had time to emerge on any Earth-like planets that might be present. Stellar ages are tabulated (with considerable uncertainty) by Schneider (footnote 1). In the sixteen cases where no published stellar age exist, we have used the mass and metallicity of the star, in conjunction with a stellar model by Mazzitelli (Underwood et al. 2003), to calculate the age at which the star would reach its present luminosity.

## 2 CLASSICAL HABITABLE ZONE AND GIANT PLANET REACH

### *2.1 Determining the classical habitable zone boundaries*

The classical habitable zone (HZ) is that range of distances from a star where water at the surface of an Earth-like planet would be in the liquid phase. Habitable zones created, for example, by tidal heating, are not included.

We have used boundaries for the HZ derived from the work of Kasting, Whitmire, & Reynolds (1993). The inner boundary is the maximum distance from the star where a runaway greenhouse effect would lead to the evaporation of all surface water, and the outer boundary is the maximum distance at which a cloud-free $CO_2$ atmosphere could maintain a surface temperature of 273K. Alternative criteria have also been applied by Kasting et al., which give boundaries to each side of these. Our choice is supported by the resulting HZ in the Solar Systems, which conforms well to what we know about Venus, the Earth, and Mars.

To obtain the HZ boundaries we need to use the stellar flux $S_b$ that occurs at each boundary. These have been established by Kasting et al. (1993, updated by Underwood et al. 2003). This flux depends mainly on $L$, but to some extent on the effective temperature $T_e$ of the star, the lower this temperature the smaller the critical flux – this is because the greater the infrared fraction the greater the heating effect on a planet. We denote the critical flux by $S_b(T_e)$, which in units of the solar constant is given by

$$S_{bri}(T_e) = 4.190\ 10^{-8}\ T_e^2 - 2.139\ 10^{-4}\ T_e + 1.296 \qquad (1a)$$

at the inner boundary (runaway greenhouse), and

$$S_{bro}(T_e) = 6.190\ 10^{-9}\ T_e^2 - 1.319\ 10^{-5}\ T_e + 0.2341 \qquad (1b)$$

at the outer boundary (maximum greenhouse), where $T_e$ is in kelvin. The boundaries are then at distances $r_i$ and $r_o$ from the star in AU given by

$$r_i = (L/S_{bri}(T_e))^{1/2} \qquad (2a)$$

$$r_o = (L/S_{bro}(T_e))^{1/2} \qquad (2b)$$

where $L$ is the luminosity of the star in solar units and $S_{bri}(T_e)$ and $S_{bro}(T_e)$ are in units of the solar constant. We have obtained $L$ and $T_e$ from measured properties of stars. $L$ (in solar units) is obtained from

$$L = 0.787 d^2\ 10^{[-0.4(V+BC)]} \qquad (3)$$

where $V$ is the apparent visual magnitude and $BC$ is the bolometric correction (the apparent bolometric magnitude is $(V + BC)$), and the distance $d$ to the star is in parsecs (pc). Values of $T_e$ are tabulated by Schneider (see Footnote 1), as are $d$ and $V$. The $BC$s are from Appendix G in Carroll and Ostlie (2007).

In the calculation of $L$ from observed quantities as in eqn(3), the uncertainty is dominated by that in $d$. Many of these distances come from Hipparcos, where the measured parallax has a median standard error of $0.97\times10^{-3}$ arcsec (Perryman et al. 1997). At 100 pc this is ±10%. From eqs(2) and (3) we see that this translates into a ±10% uncertainty in $r$. Values of $T_e$ are perhaps subject to less uncertainty. It is also the case that $S_b$ is only weakly dependent on $T_e$ (eqn(1)). For example, for our HZ boundary criteria, at around 5700 K, a change of 300 K changes $S_b$ at each boundary by only about 5%. Also, $r$ goes as the square root of $L$ and $S_b$ (eqn(2)), thus approximately halving its sensitivity to $L$ and $S_b$. The uncertainties in $L$ are thus significant but not serious.

*2.2 Determining a giant planet's gravitational reach*

With a HZ obtained in this way we need to test to what extent the planet(s) known to be present in the system prevent(s) an Earth-like planet being confined to the HZ, in the sense that the semimajor axis of the Earth-like planet does not stray outside the HZ. This is determined by the gravitational reach of the known planet(s), which in the very great majority of transiting exoplanets is a giant. The inward reach is at a distance $n_{int}R_H$ interior to the periastron of the giant planet; the outward reach is at a distance $n_{ext}R_H$ exterior to the apastron. $R_H$ is the Hill radius of a giant planet, defined by

$$R_H = \left(\frac{m_G}{3M_{star}}\right)^{1/3} a_G \qquad (4)$$

where $m_G$ is the mass of the giant planet, $a_G$ is its orbital semimajor axis, and $M_{star}$ is the mass of the star. The reaches are thus

- $a_G(1 - e_G) - n_{int}R_H$ interior to the semimajor axis of the giant's orbit
- $a_G(1 + e_G) + n_{ext}R_H$ exterior to the semimajor axis of the giant's orbit.

where $e_G$ is the eccentricity of the giant's orbit. Within these reaches confinement is unlikely. If they extend across the whole HZ then nowhere in the HZ offers a safe haven. The multipliers $n_{int}$ and $n_{ext}$ depend on the eccentricity of the giant planet's orbit and have been obtained as follows.

To obtain the $n_{int}$ and $n_{ext}$ values we earlier studied in detail seven contrasting systems known to have a giant planet, using the *MERCURY* package of orbital integrators (Chambers 1999). We could have invented systems to cover an appropriate range of orbital parameters of the giant planet. Instead, the seven systems were based on HD196050, HD216435, HD72659, HD196050, HD52265, Epsilon Eridani, and HD72659, using data from 2004. Our studies of these systems were sufficiently detailed to consume over a thousand hours of CPU time on fast PCs. Full details of how we obtained the $n_{int}$ and $n_{ext}$, are in Jones et al. (2005). The key discoveries are as follows.

- $n_{int}$ and $n_{ext}$ are sensitive only to the eccentricity $e_G$ of the giant planet's orbit (and not, for example, to $m_G/M_{star}$ and $a_G$).
- It is an increase in the eccentricity of the orbit of a small planet (Earth-like or otherwise) that leads to its ejection or collision. A giant planet can pump up this eccentricity to large values without more than a few percent change in the semimajor axis of the small planet.

The values of $n_{int}(e_G)$ and $n_{ext}(e_G)$ are shown in Figure 1, where the curves connecting the data points are cubic fits. These fit closely the $n_{int}$ and $n_{ext}$ data, with correlation coefficients $\rho$ such that $\rho^2 = 0.970$ for $n_{int}$ and 0.998 for $n_{ext}$.

Figure 1 The Hill radius multipliers $n_{int}(e_G)$ and $n_{ext}(e_G)$ for seven exoplanetary systems studied in detail, versus the eccentricity $e_G$ of the giant planet's orbit, and a cubic fit to these points. The values for the seven systems are shown as points with error bars.

The values of $n_{int} = n_{ext} = 3$ at low eccentricity are in accord with analytical values obtained, for example by Gladman (1993), for $e_G$ close to zero. Analytical solutions are not possible at higher eccentricity. See Jones et al. (2005) for a brief discussion about the relationships between $n_{int}$, $n_{ext}$, and $e_G$.

The value of $n_{ext}(e_G)$ rises particularly strongly as $e_G$ increases. It is therefore *essential*, in *any* work involving the outward gravitational reach of a planet, to include this multiplier.

Note that even if confined to the HZ, the orbital eccentricity of an Earth-like planet will generally increase due to the giant planet's gravity, and might rise to the point where the planet is carried outside the HZ for a significant fraction of its orbital period. Whether a planet could be habitable in such a case depends on the response time of the atmosphere-ocean system; Williams and Pollard (2002) conclude that a planet like the Earth probably could. For example, if the planet's eccentricity $e \sim 0.2$ the ratio of periastron stellar flux to apastron stellar flux, $[(1 + e)/(1 - e)]^2$, is about the same as the summer/winter flux ratio at mid-latitudes on Earth (due to the Earth's *obliquity*). They conclude that an 'Earth' with $e \sim 0.2$ would be habitable as long as its semimajor axis $a$ remained in the HZ. The upper limit on $e$ for an Earth-like planet to be habitable is probably between 0.5 and 0.7. For confined orbits we find from orbital integration that $e$ is usually less than about 0.3 and rarely exceeds 0.4. Greater eccentricities generally result in ejection or collision.

## 3 HABITABILITY OF THE KNOWN EXOPLANETARY SYSTEMS

### *3.1 Classification of results*

Armed with the critical distances from a giant planet, we can now see whether this reduces the extent to which the HZ would offer confinement to an Earth-like planet. There are six distinct types of configuration, labelled 1-6 in Figure 2, where the lines represent the reaches from the semimajor axis distance of the giant. Because the HZ migrates outwards as the star goes through its main-sequence lifetime, these are instantaneous configurations, and will change with stellar age. The confinement outcome relates to the configuration as follows.

| *configuration* | *confinement outcome (for Earth-like planets)* | *system habitability today* |
|---|---|---|
| 1, 2 | confinement throughout the HZ | 'Yes' |
| 3, 4, 5 | $x, y, z$ %, fraction of HZ width offering confinement | $x, y, z$ % |
| 6 | confinement nowhere in the HZ | 'No' |

Figure 2 Six configurations of the gravitational reach of a giant planet from its semimajor axis distance (black discs) with respect to the instantaneous position of the HZ. See the text for the relationship between the configuration number (1-6) and the extent to which each configuration offers confinement for Earth-like planets in the HZ, and the corresponding system habitability today.

We are interested in two scenarios.

1 The system habitability *today*, which is determined by the confinement outcome today. The possible system habitabilities today are defined above, as 'Yes', a percentage, or 'No'.

2 *Sustained* habitability, requiring the star to be at least 1.7 Gyr old. The first 0.7 Gyr covers a presumed heavy bombardment phase as on Earth, followed by at least 1 Gyr for life to emerge, which, for the Earth, is about the most pessimistic delay (Jones 2004, Chapter 3). The possible results are as follows.

- If the system habitability today is 'Yes' then the *sustained* habitability is
  – 'Yes' if the star's age is $\geq$ 1.7 Gyr and it is on the main-sequence, as indicated by luminosity class V (all stars in the transiting systems are class V)
  – 'too young' if the star's age is < 1.7 Gyr

  If the system habitability today is a % then the *sustained* habitability is
  – 'Yes' if >20%, if the star's age is $\geq$ 1.7 Gyr, and if it is on the main-sequence as indicated by luminosity class V
  – 'No' if $\leq$ 20%; the case is too marginal, so we take the pessimistic view
  – 'too young'.

- If the system habitability *today* is 'No' then the *sustained* habitability outcome is 'No'.

We have used stellar ages tabulated by Schneider (see Footnote 1). As noted in Section 1.1, for sixteen of the stars no age is tabulated, and so we have obtained plausible estimates from the stellar evolution model of Mazzitelli (Underwood et al. 2003).

Note that to establish sustained habitability we require an estimate of how much the HZ has moved outwards in the past 1000 Ma. Except in very marginal cases this can be done well enough from the star's mass for class V stars. Marginal cases are excluded by the ≤ 20% criterion given above. NB "sustained habitability" does *not* imply habitability throughout the main sequence lifetime of the star. (To examine this possibility a stellar evolution model would be required to establish how the HZ moves outwards from star birth to the start of its transition to becoming a giant star; see Jones et al. 2005).

We initially assume that a giant planet interior to the HZ will *not* have ruled out the presence of Earth-like planet beyond the giant, even though the giant will probably have got there by traversing the HZ (Chambers 2009). Subsequently, we examine the result when the formation of Earth-like planets *is* ruled out by such traversal.

### *3.2 Results*

Table 3 summarises the results of our analysis applied to the 79 transiting exoplanetary systems in Schneider as at April 2010. Note the following.

1 The systems are ordered by increasing orbital period of the planet closest to each star, which ranges from 0.78884 days for WASP-19 b, to 111.4364 days for HD 80606 b. **(**Only **two** other (innermost) planets have periods exceeding 10 days, HD17156 b and CoRot-9 b.)

2 In 3 cases Schneider does not give the mass of the host star. As these are all G**V** stars we have given them the solar mass; shown in *italics* in the $M_{\text{star}}$ column.

3 In 19 cases either the stellar distance *d* or/and the V magnitude is/are not tabulated by Schneider, then, in place of calculating the stellar luminosity $L_{\text{star}}$ from eqn(3), the luminosity is obtained from the spectral class, as tabulated in Appendix G of Carroll and Ostlie (2007). This is also the procedure in a further two cases, SWEEPS-11 and SWEEPS-04, otherwise the luminosity values are far too low, perhaps because the distances *d* are uncertain. The 21 $L_{\text{star}}$ values are shown in *italics* in the $L_{\text{star}}$ column.

4 In the 16 cases where no published stellar age exist, we have used the mass and metallicity of the star, in conjunction with a stellar model by Mazzitelli (Underwood et al. (2003); shown asterisked in the age column, and, also asterisked in the age uncertainly column where it is possible to estimate a value.)

5 In 12 cases Schneider gives neither stellar age nor metallicity of the star. An age of 1.7 Gyr is assumed, marked with † in the age column with an age uncertainty NA (not available)

6 In 6 cases no value for orbital eccentricity is given. An asterisked value of 0 is assumed.

7 For the 3 systems with more than one giant planet, the $nR_H$ for each giant is obtained and the confinement outcome is based on the combined gravitational reaches. The configuration is for the giant with the worst (highest numbered) configuration.

TABLE 3

DATA ON THE KNOWN TRANSITING EXOPLANETARY SYSTEMS AND THEIR HOST STAR

{SEE END OF DOCUMENT}

Table 3 lists neither the configurations nor the associated habitabilities of the transiting exoplanetary systems. This is because, as we anticipated, with the giant planets predominantly in low eccentricity orbits, considerably closer to the star than the inner boundary of the HZ, the predominant outcome is

- configuration, 2
- system habitability today, ‘Yes’.

There are just two exceptions, both with configuration 6 and ‘No’ for system habitability today:

- HAT-P-13, a 1.22 solar mass G4V star, with two planets. Planet b has a mass 0.851 $m_J$, and a near-circular orbit with $a$ = 0.0426 AU. Planet c (unconfirmed) has a mass 15.2 $m_J$ , with $a$ = 1.186 AU. If planet c is confirmed it will be responsible for making this system uninhabitable.
- HD80606, a 0.9 solar mass G5V star with one known planet, with a mass 3.94 $m_J$, and a highly eccentric orbit, with $e$ = 0.9337 and $a$ = 0.449 AU. It is the high eccentricity that makes this system uninhabitable.

Note that no semimajor axis is given for OGLE2-TR-L9, so there is no evaluation of habitability.

Sustained habitability outcomes (SHOs) are less easily summarized. You saw above that an essential requirement for ‘Yes’ is that the star’s age is ≥ 1.7 Gyr and that it is on the main-sequence. We have obtained the SHO for the best estimate of the age of the star, and for a minimum and maximum age obtained from the age uncertainty. If no such uncertainly is available then only the SHO for the best estimate has been obtained. In addition we have eased the age requirement from 1.7 Gyr to 1.0 Gyr and obtained three more SHOs per system. With the exceptions of HAT-P-13 and HD 80606 (noted earlier) the SHO outcome is either 'Yes' or 'too young'. There are 76 such cases. For the 1.7 Gyr minimum age for sustained habitability, the number of stars that are too young ranges from 40 at minimum stellar age, to 28 at the best age estimate, to 24 at maximum stellar age. For the 1.0 Gyr minimum age for sustained habitability, the numbers are respectively 28, 22, and 16. These trends with stellar age are, of course, in line with expectations.

Table 4 lists the details for a small sample of the known transiting exoplanetary systems. As well as HAT-P-13 and HD 80606, a few typical examples are also shown. The main purpose of this Table is to show the type of details that will be more varied when many more transiting exoplanetary systems have been discovered.

TABLE 4

SUSTAINED HABITABILTY OUTCOMES (SHO) FOR A SMALL SAMPLE OF THE KNOWN TRANSITING EXOPLANETARY SYSTEMS

{SEE END OF DOCUMENT}

## 4 DISCUSSION OF RESULTS

The stellar age is an important determinant of habitability in the transiting exoplanetary systems, as is the stability of orbits of small planets in the HZ. But before we conclude that a substantial proportion of the known transiting exoplanetary systems offer sustained habitability, and that almost all offer habitability today, we must ask the question "could Earth-like planets have formed in the HZ?"

It is widely believed that giant planets close to their stars could not have formed *in situ*, because there would not have been enough suitable nebular material. In particular, only beyond several AU from a young star would the abundant water have been able to condense and help to build massive icy-rocky cores that could capture copious quantities of gaseous hydrogen and helium from the nebula (Encrenaz 2007). So if giant planets could only form beyond several AU, those we find at small fractions of an AU must have moved there. There is a well-established scheme for such migration to have occurred. In brief, the growing giant planet interacts gravitationally with the remaining nebular gases such that the giant planet spirals inwards, until halted by one or more of a variety of mechanisms; see, for example, Chambers (2009) for further details.

To get to its final small orbit the growing giant has to cross the HZ. At this time the HZ would be full of dust, planetesimals, and embryos, with the potential to form one or more Earth-like planets.

But if the HZ-crossing giant removes this material by scattering or capture, so that the HZ is utterly depleted, then no Earth-like planets could form after migration. In this case not a single one of the known transiting exoplanetary systems will offer habitability.

Fortunately, simulations have shown that, even though a migrating giant does remove beyond use a large proportion of the dust, planetesimals, and embryos, sufficient will remain in many cases to form an Earth-like planet, particularly of lower mass. Mandell and Sigurdsson (2003) have shown that even when the entire HZ is traversed by a giant planet, a significant fraction of any pre-formed terrestrial planets could survive, eventually returning to circular orbits fairly close to their original positions. Fogg and Nelson (2005) have shown that post-migration formation of Earth-like planets from planetesimals and planetary embryos is likely. This is a more optimistic result than that of Armitage (2003), who concluded that post-migration formation of Earth-like planets might be unlikely, though he concentrated on the effect of giant migration on planet-forming dust rather than on planetesimals and planetary embryos. This problem needs further study.

It is thus not unlikely that low mass Earth-like planets await discovery in the HZ of a significant proportion of transiting exoplanetary systems, even in those presently known tagged “yes” for sustained habitability.

Though there have been previous studies of the habitability of large numbers of known exoplanetary systems by us and others, for example Sándor et al. (2007), Jones et al. (2006), Menou and Tabachnik (2003), Turnbull and Tarter (2003), none has focussed on transiting systems. In any case, the great majority of transiting systems have been discovered in the last few years. We know of no other assessments of the habitability of such systems.

## 5 CONCLUSIONS AND FUTURE WORK

Our main conclusion is that low mass Earth-like planets are likely awaiting discovery in the HZ of a significant proportion of transiting exoplanetary systems, provided that Earth-like planets could have formed after a giant planet migrated through the HZ. This latter requirement is so important that, though work to date is encouraging, further work needs to be done on such post-migration formation. Such work is not only relevant to transiting systems, but to all exoplanetary systems with a giant planet in a orbit closer to the star than the inner boundary of the HZ

This paper has demonstrated a methodology for establishing the habitability of transiting systems, just as our earlier work (Jones et al. 2005, 2006) demonstrated its applicability to all exoplanetary systems. We await further discoveries of transiting systems, and further exploration of those already known, to update our work, in the expectation that the habitability outcomes will show a greater variety when planets further from their star are discovered. There are many ground-based searches for transiting planets, and recently the spacecraft Kepler has joined the hunt (http://kepler.nasa.gov/).

We are grateful to John Chambers for discussions, and to a reviewer for helpful comments.

FIGURE CAPTIONS

Fig. 1.–The Hill radius multipliers $n_{int}(e_G)$ and $n_{ext}(e_G)$ for seven exoplanetary systems studied in detail, versus the eccentricity $e_G$ of the giant planet's orbit, and a cubic fit to these points. The values for the seven systems are shown as points with error bars.

Fig. 2.–Six configurations of the gravitational reach of a giant planet from its semimajor axis distance (black discs) with respect to the instantaneous position of the HZ. See the text for the relationship between the configuration number (1-6) and the extent to which each configuration offers confinement for Earth-like planets in the HZ, and the corresponding system habitability today.

## FIGURE 1

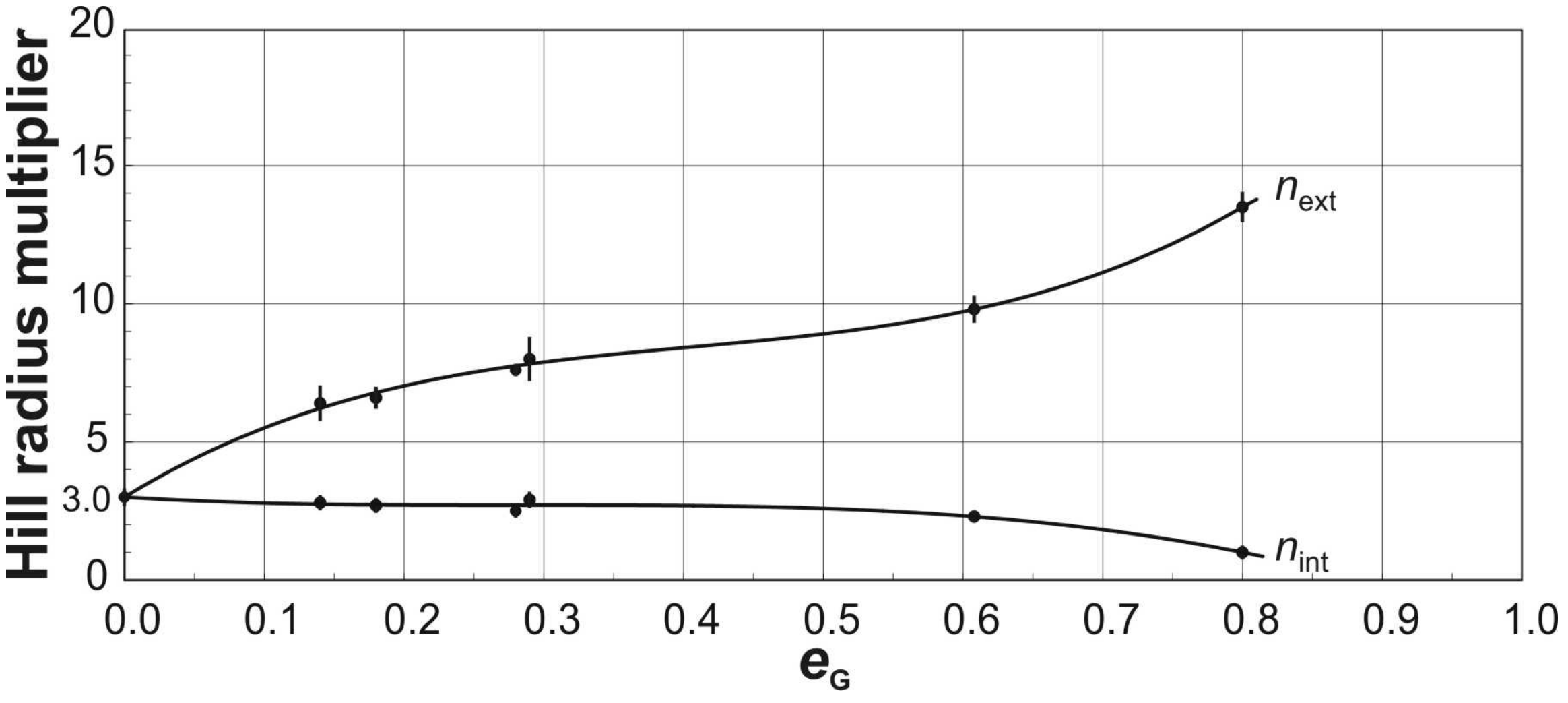

## FIGURE 2

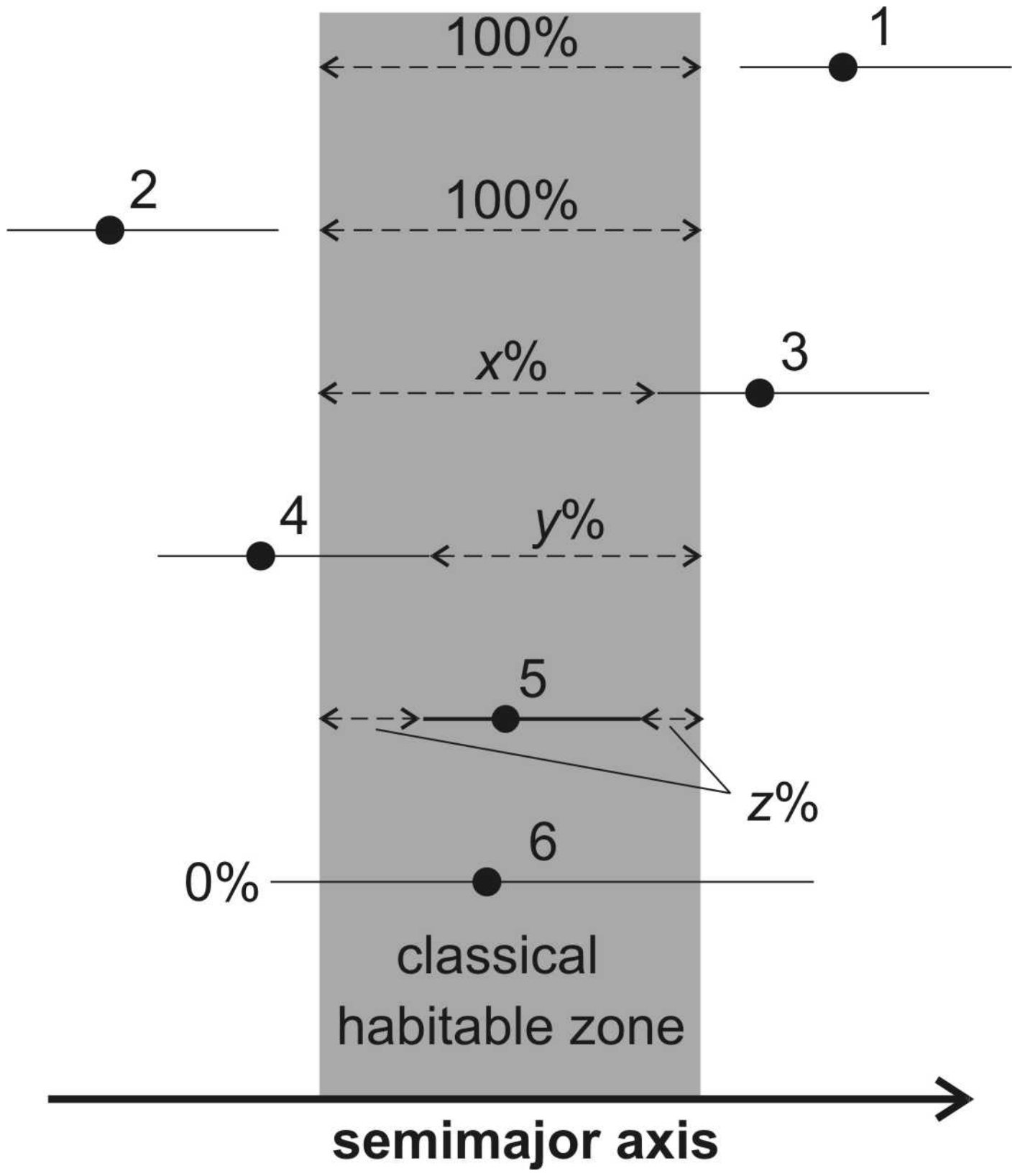

# Table 3 Data on the known transiting exoplanetary systems and their host star
## (ordered by period of (innermost) planet – period(s) not shown)

| | Properties of the stars | | | | | | | | | | | Properties of the planets | | | | |
|---|---|---|---|---|---|---|---|---|---|---|---|---|---|---|---|---|
| Star, planet(s) | $M_{star}$ /$M_{Sun}$ | *d*/pc | Spec + lum class | V | BC | $L_{star}$ /$L_{Sun}$ | $T_{eff}$ /K | Age /Gyr | Age± /Gyr | $r_i$ /AU | $r_o$ /AU | $m/m_J$ | $r/r_J$ | *a*/AU | *e* | *i*/° |
| WASP-19 b | 0.95 | | G8V | 12.3 | -0.4 | *0.66* | 5500 | 0.6 | −0.04+0.05 | 0.70 | 1.38 | 1.15 | 1.31 | 0.0164 | 0.02 | 80.8 |
| CoRoT-7 b | 0.93 | 150 | K0V | 11.7 | -0.31 | 0.49 | 5275 | 1.5 | −0.3+0.8 | 0.61 | 1.21 | 0.0151 | 0.15 | 0.0172 | 0 | 80.1 |
| c | | | | | | | | | | | | 0.0264 | | 0.046 | 0 | |
| WASP-18 b | 1.281 | 100 | F9V | 9.3 | -0.17 | 1.75 | 6400 | 0.63 | −0.53+0.95 | 1.04 | 2.09 | 10.43 | 1.165 | 0.02047 | 0.0092 | 86 |
| WASP-12 b | 1.35 | 267 | G0V | 11.69 | -0.18 | 1.40 | 6300 | 0.9* | <1 | 0.94 | 1.88 | 1.41 | 1.79 | 0.0229 | 0.049 | 83.1 |
| OGLE-TR-56 b | 1.17 | 1500 | G2V | 16.6 | -0.2 | 0.49 | 5790 | 2 | >age | 0.58 | 1.16 | 1.29 | 1.3 | 0.0225 | 0 | 78.8 |
| WASP-33 b | 1.495 | 116 | A5V | 8.3 | -0.15 | 5.82 | 7400 | ~0* | ~0 | 1.71 | 3.50 | 4.10 | 1.497 | 0.02555 | 0* | 87.67 |
| TrES-3 | 0.924 | | G2V | 12.4 | -0.2 | *1.00* | 5720 | 3.12* | | 0.84 | 1.66 | 1.92 | 1.295 | 0.0226 | 0 | 82.15 |
| WASP-4 b | 0.9 | 300 | G7V | 12.6 | -0.4 | 0.93 | 5500 | 1.7† | NA | 0.83 | 1.64 | 1.1215 | 1.416 | 0.023 | 0 | 89.35 |
| OGLE-TR-113 b | 0.78 | 1500 | K4V/KIV | | -0.55 | *0.46* | 4990 | 0.7 | >age | 0.61 | 1.19 | 1.32 | 1.09 | 0.0229 | 0 | 89.4 |
| CoRoT-1 b | 0.95 | 460 | G0V | 13.6 | -0.18 | 0.71 | 5950 | 0.99* | ~1 | 0.69 | 1.38 | 1.03 | 1.49 | 0.0254 | 0 | 85.1 |
| GJ 1214 b | 0.157 | 13 | M6 | 14.67 | -3.21 | 0.00 | 3026 | 6 | −3+4 | 0.06 | 0.12 | 0.0179 | 0.2415 | 0.014 | 0.27 | 88.62 |
| WASP-5 b | 1.021 | 297 | G4V | 12.26 | -0.2 | 1.04 | 5880 | 3 | ± 1.4 | 0.84 | 1.68 | 1.637 | 1.171 | 0.02729 | 0 | 85.8 |
| OGLE-TR-132 b | 1.26 | 1500 | F8V/F7V | | -0.16 | *2.00* | 6210 | 0.9* | <1 | 1.13 | 2.26 | 1.14 | 1.18 | 0.0306 | 0 | 85 |
| CoRoT-2 b | 0.97 | 300 | K0V | 12.57 | -0.31 | 0.88 | 5625 | 1.7† | NA | 0.80 | 1.58 | 3.31 | 1.465 | 0.0281 | 0 | 87.84 |
| SWEEPS-11 | 1.1 | ~2000 | G0V | 19.83 | -0.18 | *1.25* | 5940 | 1.7† | NA | 0.92 | 1.83 | 9.7 | 1.13 | 0.03 | 0* | 84 |
| WASP-3 b | 1.24 | 223 | F7V | 10.64 | -0.16 | 2.52 | 6400 | 1.7† | NA | 1.25 | 2.50 | 1.76 | 1.31 | 0.0317 | 0 | 85.06 |
| WASP-2 b | 0.84 | 144 | K1V | 11.98 | -0.37 | 0.37 | 5200 | 1.7† | NA | 0.54 | 1.05 | 0.914 | 1.017 | 0.03138 | 0 | 84.8 |
| HAT-P-7 b | 1.47 | 320 | F8V | 10.5 | -0.16 | 5.89 | 6350 | 0.0085* | | 1.92 | 3.84 | 1.8 | 1.421 | 0.0379 | 0 | 84.1 |
| HD 189733 b | 0.8 | 19.3 | K1-K2V | 7.67 | -0.43 | 0.37 | 4980 | 0.6 | >age | 0.55 | 1.08 | 1.13 | 1.138 | 0.03099 | 0 | 85.76 |
| WASP-14 b | 1.319 | 160 | F5V | 9.75 | -0.14 | 2.89 | 6475 | 0.75 | ± 0.25 | 1.33 | 2.66 | 7.725 | 1.259 | 0.037 | 0.0903 | 84.79 |
| WASP-24 b | 1.129 | 330 | F8-9V | 11.3 | -0.16 | 3.00 | 6075 | 1.6 | -1.6+2.1 | 1.41 | 2.80 | 1.032 | 1.104 | 0.0359 | 0 | 85.71 |
| TrES-2 | 0.98 | 220 | G0V | 11.41 | -0.18 | 1.23 | 5850 | 5.1 | ± 2.7 | 0.92 | 1.82 | 1.199 | 1.272 | 0.03556 | 0 | 83.62 |
| OGLE2-TR-L9 b | 1.52 | 900 | F3V | | -0.11 | *3.20* | 6933 | 0.00066 | | 1.33 | 2.70 | 4.5 | 1.61 | ? | 0* | |
| WASP-1 b | 1.24 | | F7V | 11.79 | -0.16 | *2.30* | 6200 | 1.7† | NA | 1.22 | 2.43 | 0.89 | 1.358 | 0.0382 | 0 | 83.9 |
| XO-2 b | 0.98 | 149 | K0V | 11.18 | -0.31 | 0.78 | 5340 | 2 | | 0.77 | 1.52 | 0.57 | 0.973 | 0.0369 | 0 | 88.58 |
| GJ 436 b | 0.452 | 10.2 | M2.5V | 10.68 | -2 | 0.03 | 3684 | 6 | −5+4 | 0.16 | 0.32 | 0.072 | 0.438 | 0.02872 | 0.15 | 85.8 |
| WASP-26 b | 1.12 | 250 | G0V | 11.3 | -0.18 | 1.75 | 5950 | 6 | ± 2 | 1.09 | 2.16 | 1.02 | 1.32 | 0.04 | 0 | 82.5 |
| HAT-P-5 b | 1.16 | 340 | G0V | 12 | -0.18 | 1.70 | 5960 | 2.6 | ± 1.8 | 1.07 | 2.13 | 1.06 | 1.26 | 0.04075 | 0 | 86.75 |
| HD 149026 b | 1.3 | 78.9 | G0IV | 8.15 | -0.18 | 3.18 | 6147 | 2 | ± 0.8 | 1.44 | 2.87 | 0.359 | 0.654 | 0.04313 | 0 | 85.3 |
| HAT-P-3 b | 0.936 | 140 | K0V | 11.86 | -0.31 | 0.37 | 5185 | 0.4 | −0.3+6.5 | 0.54 | 1.06 | 0.599 | 0.89 | 0.03894 | 0 | 87.24 |
| HAT-P-13 b | 1.22 | 214 | G4V | 10.62 | -0.21 | 2.47 | 5638 | 5 | −0.8+2.5 | 1.33 | 2.63 | 0.851 | 1.28 | 0.0426 | 0.021 | 83.4 |
| c | | | | | | | | | | | | 15.2 | | 1.186 | 0.691 | |
| TrES-1 | 0.87 | 157 | K0V | 11.79 | -0.31 | 0.50 | 5150 | 2.5 | ±1.4 | 0.62 | 1.23 | 0.61 | 1.081 | 0.0393 | 0 | 88.4 |
| HAT-P-4 b | 1.26 | 310 | F8V | 11.2 | -0.16 | 2.90 | 5860 | 4.2 | −0.6+2.6 | 1.41 | 2.80 | 0.68 | 1.27 | 0.0446 | 0 | 89.9 |
| HAT-P-8 b | 1.28 | 230 | F8V | 10.17 | -0.16 | 4.13 | 6200 | 3.4 | ± 1 | 1.63 | 3.25 | 1.52 | 1.5 | 0.0487 | 0 | 87.5 |
| WASP-10 b | 0.71 | 90 | K5V | 12.7 | -0.72 | 0.10 | 4675 | 0.8 | ± 0.2 | 0.29 | 0.58 | 3.06 | 1.08 | 0.0371 | 0.057 | 86.8 |
| OGLE-TR-10 b | 1.18 | 1500 | F8V | | -0.16 | *1.70* | 6250 | 1.1 | >age | 1.04 | 2.08 | 0.63 | 1.26 | 0.04162 | 0 | 84.5 |
| WASP-16 b | 1.022 | | G3V | 11.3 | -0.21 | *1.00* | 5550 | 2.3 | −2.2+5.8 | 0.85 | 1.69 | 0.855 | 1.008 | 0.0421 | 0 | 85.22 |
| XO-3 b | 1.213 | 260 | F5V | 9.8 | -0.14 | 7.28 | 6429 | 2.82 | −0.82+0.58 | 2.12 | 4.24 | 11.79 | 1.217 | 0.0454 | 0.26 | 84.2 |
| HAT-P-12 b | 0.73 | 142.5 | K3V | 12.84 | -0.5 | 0.19 | 4650 | 2.5 | ± 2 | 0.40 | 0.78 | 0.211 | 0.959 | 0.0384 | 0 | |
| Kepler 4 b | 1.223 | 550 | G0V | 12.7 | -0.18 | 2.34 | 5857 | 4.5 | ± 1.5 | 1.27 | 2.52 | 0.077 | 0.357 | 0.0456 | 0 | 89.76 |
| Kepler 6 b | 1.209 | | F8V | | -0.16 | *1.68* | 5647 | 3.8 | ± 1 | 1.10 | 2.17 | 0.669 | 1.323 | 0.04567 | 0 | 86.8 |
| WASP-6 b | *1* | 307 | G8V | 12.4 | -0.4 | 1.18 | 5310 | 1.7† | NA | 0.95 | 1.86 | 0.503 | 1.224 | 0.0421 | 0.054 | 88.47 |
| WASP-28 b | 1.08 | 334 | F8V-G0V | 12 | -0.17 | 1.63 | 6100 | 1.35* | | 1.03 | 2.06 | 0.91 | 1.12 | 0.0455 | 0.046 | 89.1 |
| Kepter-8 b | 1.213 | 1330 | F8V | 13.9 | -0.16 | 4.44 | 6213 | 3.84 | ± 1.5 | 1.69 | 3.37 | 0.603 | 1.419 | 0.0483 | 0 | 84.07 |
| HD 209458 b | 1.01 | 47 | G0V | 7.65 | -0.18 | 1.79 | 5942 | 4 | ± 2 | 1.10 | 2.19 | 0.685 | 1.32 | 0.04707 | 0.07 | 86.677 |
| WASP-22 b | 1.1 | 300 | G0V | 12 | -0.18 | 1.33 | 6000 | 0.6* | | 0.94 | 1.87 | 0.56 | 1.12 | 0.0468 | 0.023 | 89.2 |
| Kepler-5 b | 1.374 | | F5V | | -0.14 | *2.56* | 6297 | 0.9* | <1 | 1.27 | 2.54 | 2.114 | 1.431 | 0.05064 | 0 | 86.3 |
| TrES-4 | 1.384 | 440 | F8V | 11.592 | -0.16 | 4.07 | 6100 | 4.7 | ± 2 | 1.64 | 3.26 | 0.919 | 1.799 | 0.05091 | 0 | 82.86 |
| OGLE-TR-211 b | 1.33 | | F8V | | -0.16 | *1.70* | 6325 | 0.9* | <1 | 1.03 | 2.07 | 1.03 | 1.36 | 0.051 | 0 | 87.2 |
| WASP-11/HAT-P-10 b | 0.82 | 125 | K3V | 11.89 | -0.5 | 0.34 | 4980 | 11.2 | ± 4.1 | 0.52 | 1.03 | 0.46 | 1.045 | 0.0439 | 0 | 88.5 |
| WASP-17 b | 1.2 | | F6V | 11.6 | -0.14 | *2.50* | 6550 | 3 | −2.6+0.9 | 1.23 | 2.46 | 0.49 | 1.74 | 0.051 | 0.129 | 87.8 |
| WASP-15 b | 1.18 | 308 | F5V | 10.9 | -0.14 | 3.71 | 6300 | 3.9 | −1.3+2.8 | 1.53 | 3.06 | 0.542 | 1.428 | 0.0499 | 0 | 85.5 |
| WASP-25 b | 1 | 169 | G4V | 11.9 | -0.21 | 0.47 | 5750 | 0.9* | <1 | 0.58 | 1.14 | 0.58 | 1.26 | 0.0474 | 0 | 87.7 |
| HAT-P-6 b | 1.29 | 200 | F5V | 10.5 | -0.14 | 2.26 | 6570 | 2.3 | −0.7+0.5 | 1.16 | 2.33 | 1.057 | 1.33 | 0.05235 | 0 | 85.51 |
| Lupus-TR-3 b | 0.87 | | K1V | 17.4 | -0.37 | *0.55* | 5000 | 1.7† | NA | 0.66 | 1.31 | 0.81 | 0.89 | 0.0464 | 0 | 88.3 |
| HAT-P-9 b | 1.28 | 480 | F5V/F7V | | -0.14 | *2.00* | 6350 | 1.6 | −1.4+1.8 | 1.12 | 2.24 | 0.78 | 1.4 | 0.053 | 0 | 86.5 |
| WASP-29 b | 0.824 | 80 | K4V | 11.3 | -0.55 | 0.25 | 4800 | 1* | ~1 | 0.46 | 0.90 | 0.248 | 0.74 | 0.0456 | 0 | 87.96 |
| XO-1 b | 1 | 200 | G1V | 11.3 | -0.19 | 1.13 | 6000 | 4.5 | ± 2 | 0.87 | 1.73 | 0.9 | 1.184 | 0.0488 | 0 | 89.31 |
| OGLE-TR-182 b | 1.14 | | G0V | 16.84 | -0.18 | *1.50* | 5924 | 0.99* | ~1 | 1.01 | 2.00 | 1.01 | 1.13 | 0.051 | 0 | 85.7 |
| OGLE-TR-111 b | 0.82 | 1500 | K0V | | -0.31 | *0.55* | 5150 | 1.1 | >age | 0.66 | 1.29 | 0.53 | 1.067 | 0.047 | 0 | 88.1 |
| CoRoT-5 b | 1 | 400 | F9V | 14 | -0.16 | 0.37 | 6100 | 6.9 | ± 1.4 | 0.49 | 0.98 | 0.467 | 1.388 | 0.04947 | 0.09 | 85.83 |
| XO-4 b | 1.32 | 293 | F5V | 10.7 | -0.14 | 4.03 | 5700 | 2.1 | ± 0.6 | 1.69 | 3.35 | 1.72 | 1.34 | 0.0555 | 0 | 88.7 |
| XO-5 b | 0.88 | 255 | G8V | 12.13 | -0.4 | 1.04 | 5510 | 8.5 | ± 0.8 | 0.87 | 1.73 | 1.077 | 1.089 | 0.0487 | 0 | 86.8 |
| SWEEPS-04 | 1.24 | ~2000 | F8V | 18.8 | -0.16 | *1.70* | 6200 | 1.7† | NA | 1.05 | 2.09 | 3.8 | 0.81 | 0.055 | 0* | 87 |
| CoRoT-3 b | 1.37 | 680 | F3V | 13.3 | -0.11 | 1.93 | 6740 | 2 | −0.4+0.8 | 1.06 | 2.13 | 21.66 | 1.01 | 0.057 | 0 | 85.9 |
| WASP-21 b | 1.01 | 230 | G3V | 11.6 | -0.21 | 1.16 | 5800 | 3.13* | | 0.90 | 1.78 | 0.3 | 1.07 | 0.052 | 0 | 88.75 |
| WASP-13 b | *1* | 156 | G1V | 10.42 | -0.18 | 1.54 | 5826 | 1.7† | NA | 1.03 | 2.04 | 0.46 | 1.21 | 0.0527 | 0 | 86.9 |
| HAT-P-1 b | 1.133 | 139 | G0V | 10.4 | -0.18 | 1.24 | 5975 | 3.6 | | 0.91 | 1.82 | 0.524 | 1.225 | 0.0553 | 0.067 | 86.28 |
| HAT-P-14 b | 1.386 | 205 | F3V | 9.98 | -0.12 | 3.76 | 6600 | 1.3 | ± 0.4 | 1.50 | 3.00 | 2.232 | 1.15 | 0.0606 | 0.107 | 83.5 |
| Kepler-7 b | 1.347 | | F6V | | -0.15 | *2.20* | 5933 | 3.5 | ± 1 | 1.22 | 2.43 | 0.433 | 1.478 | 0.06224 | 0 | 86.5 |
| HAT-P-11 b | 0.81 | 38 | K4V | 9.59 | -0.55 | 0.28 | 4780 | 6.5 | −4.1+5.9 | 0.48 | 0.94 | 0.081 | 0.452 | 0.053 | 0.198 | 88.5 |
| WASP-7 b | 1.28 | 140 | F5V | 9.51 | -0.14 | 2.76 | 6400 | 1.7† | NA | 1.31 | 2.61 | 0.96 | 0.915 | 0.0618 | 0 | 89.6 |
| HAT-P-2 b | 1.36 | 118 | F8V | 8.71 | -0.16 | 4.17 | 6290 | 2.7 | ± 0.5 | 1.62 | 3.24 | 9.09 | 1.157 | 0.06878 | 0.5171 | 86.72 |
| WASP-8 b | *1* | 49 | G6V | 9.9 | -0.3 | 0.27 | 5600 | 1.7† | NA | 0.44 | 0.88 | 2.23 | 1.17 | 0.0793 | 0* | |
| CoRoT-6 b | 1.055 | | F5V | 13.9 | -0.14 | *2.60* | 6090 | 2.35* | | 1.31 | 2.60 | 2.96 | 1.166 | 0.0855 | 0.1 | 89.07 |
| CoRoT-4 b | 1.1 | | F0V | 13.7 | -0.09 | *5.20* | 6190 | 1 | −0.3+1 | 1.83 | 3.65 | 0.72 | 1.19 | 0.09 | 0 | 90 |
| HD 17156 b | 1.24 | 78.24 | G0V | 8.17 | -0.18 | 3.07 | 6079 | 3.06 | −0.76+0.64 | 1.42 | 2.83 | 3.212 | 1.023 | 0.1623 | 0.6753 | 86.2 |
| c | | | | | | | | | | | | 0.063 | | 0.481 | 0.136 | |
| CoRot-9 b | 0.99 | 460 | G3V | 13.7 | -0.21 | 0.67 | 5625 | 0.9* | <1 | 0.69 | 1.37 | 0.84 | 1.05 | 0.407 | 0.11 | 89.9 |
| HD 80606 b | 0.9 | 58.38 | G5V | 8.93 | -0.3 | 0.95 | 5370 | 7.63 | | 0.84 | 1.66 | 3.94 | 1.029 | 0.449 | 0.9337 | 89.285 |

| | |
|---|---|
| Mass of star: | In three cases, no value given in Schneider. As these are all G stars a solar mass is assumed. Shown in *italics*. |
| Luminosity of star: | If either the stellar distance or/and the V magnitude is/are unknown, then, in place of using eqn(3), the luminosity is calculated from the stellar mass. Shown in *italics*. |
| Age of star: | In **16** cases no value is given in Schneider. In these cases the age has been calculated from the stellar mass and metallicity using the stellar evolution model Mazzitelli. Asterisked * In 12 cases neither stellar age nor metallicity is given in Schneider. An age of 1.7 Gyr is assumed. Marked by †. The corresponding age uncertainty is shown as "NA", not available, because there is no well-defined uncertainty in the age calculated from Mazzitelli. |
| Orbital eccentricity of planet: | In **6** cases no value is given in Schneider. A value of 0 is assumed. Asterisked*. |
| SWEEPS-11, SWEEPS-04: | The luminosity has been obtained as in "Luminosity of star", otherwise the values are far too low, perhaps because the distances *d* are uncertain. |

## Table 4 Sustained habitability outcomes (SHOs) for a small sample of the known transiting exoplanetary systems

| | Minimum age for sustained habitability 1.7 Gyr | | | | | | Minimum age for sustained habitability 1.0 Gyr | | | | | |
|---|---|---|---|---|---|---|---|---|---|---|---|---|
| **Star, planet** | **Age/Gyr** | **SHO** | **Age/Gyr** | **SHO** | **Age /Gyr** | **SHO** | **Age/Gyr** | **SHO** | **Age/Gyr** | **SHO** | **Age/Gyr** | **SHO** |
| WASP-19 b | 5.60E+08 | too young | 6.00E+08 | too young | 6.50E+08 | too young | 5.60E+08 | too young | 6.00E+08 | too young | 6.50E+08 | too young |
| CoRoT-7 b, c | 1.20E+09 | too young | 1.50E+09 | Yes | 2.30E+09 | Yes | 1.20E+09 | Yes | 1.50E+09 | Yes | 2.30E+09 | Yes |
| OGLE-TR-56 b | 2.00E+09 | Yes | 2.00E+09 | Yes | 2.00E+09 | Yes | 2.00E+09 | Yes | 2.00E+09 | Yes | 2.00E+09 | Yes |
| HAT-P-13 b, c | 4.20E+09 | No | 5.00E+09 | No | 7.50E+09 | No | 4.20E+09 | No | 5.00E+09 | No | 7.50E+09 | No |
| Kepler-5 b | 9E+08 | too young | 9E+08 | too young | 9E+08 | too young | 9E+08 | too young | 9E+08 | too young | 9E+08 | too young |
| Lupus-TR-3 b | 1.70E+09 | Yes | 1.70E+09 | Yes | 1.70E+09 | Yes | 1.70E+09 | Yes | 1.70E+09 | Yes | 1.70E+09 | Yes |
| HAT-P-9 b | 2.00E+08 | too young | 1.60E+09 | too young | 3.40E+09 | Yes | 2.00E+08 | too young | 1.60E+09 | Yes | 3.40E+09 | Yes |
| HD 17156 b, c | 2.30E+09 | Yes | 3.06E+09 | Yes | 3.70E+09 | Yes | 2.30E+09 | Yes | 3.06E+09 | Yes | 3.70E+09 | Yes |
| HD 80606 b | 7.63E+09 | No | 7.63E+09 | No | 7.63E+09 | No | 7.63E+09 | No | 7.63E+09 | No | 7.63E+09 | No |

Based on Mazzitelli stellar modelling